\documentclass[twocolumn,floatfix]{revtex4-1}

\usepackage{graphicx}
\usepackage[ansinew]{inputenc}
\usepackage{color}

\begin{document}

\title{Step velocity tuning of SrRuO$_{3}$ step flow growth on SrTiO$_{3}$ }

\author{D. Est\`eve}
\author{T. Maroutian}
\author{V. Pillard}
\author{Ph. Lecoeur}
\affiliation{Institut d'Electronique Fondamentale, Univ Paris-Sud, CNRS UMR 8622, F-91405 Orsay Cedex, France}

\begin{abstract}

Taking advantage of vicinal (001) SrTiO$_{3}$ substrates with different mean terrace
widths, the heteroepitaxial growth of SrRuO$_{3}$ in the step flow mode has been
mapped as a function of mean step velocity. Transition between stable and
unstable step flow is shown to occur at a well-defined critical step velocity,
with a step bunching instability observed below this threshold. The ability to pass
from unstable to stable step flow during growth upon increasing the mean step
velocity is demonstrated. This result is discussed in terms of a stress-based
driving force for step bunching in competition with an effective step-up adatom
current.

\end{abstract}

\maketitle

When considering the use of complex oxides in solid state electronic devices, a core issue
is the control of the interfaces regarding both their structure and chemical nature. Sharpness
of the interfaces at single-layer level is for example a requisite for well-defined barriers
in magnetic tunnel junctions \cite{mtj09}. Similar high-end control of atomic termination and
oxidation state between heteroepitaxial oxide films allows for the obtention of a confined
electron gas at the interface \cite{hwa06,lao08}. First step towards optimal interface quality is to start
from a substrate with a homogeneous surface termination, like the single-terminated (001) SrTiO$_{3}$
substrate exhibiting only TiO$_{2}$ planes after suitable preparation \cite{kaw94,bla98}. Next, a
two-dimensional (2D) growth mode will often be preferred to limit the development of surface
roughness. In this way, the pulsed Laser deposition (PLD) technique is able to build on the
quality of the substrate surface to obtain fully strained films with the target stoichiometry, as
has been shown in particular for SrRuO$_{3}$ growth on SrTiO$_{3}$ \cite{bla04,fon04}.

Going the 2D growth route, an appealing scheme is to start from a vicinal surface and
to control both the atomic step positions and surface termination during growth in the so-called
step flow mode \cite{sflow}. The latter is usually achieved through proper tuning of deposition
temperature ($T$) and flux ($F$) with respect to the substrate mean terrace width ($L$). In
this work, we studied the heteroepitaxy of SrRuO$_{3}$ on TiO$_{2}$-terminated SrTiO$_{3}$ substrate as a model
system, for which step flow growth is readily achieved \cite{bla04} while being found unstable against
step bunching \cite{zha05,zha07}. In Ref. \onlinecite{zha05} and \onlinecite{zha07}, the origin of this
instability has been traced back to a strain-driven attractive interaction between steps \cite{ter95}, with
a model being proposed to account for the transition between stable and unstable step flow.
However, a comprehensive experimental study of this transition as a function of growth parameters is found
lacking, in particular going from unstable to stable step flow. Focusing on $F$ as tunable growth parameter,
we show experimentally that the mean step velocity $V = FL$ of the flowing step train determines the stability
of this growth mode for a strained layer, with a reversible transition evidenced at a critical velocity value.
For a fixed temperature, we mapped SrRuO$_{3}$ growth as a function of mean step velocity, by varying the
deposition flux and using (001) SrTiO$_{3}$ vicinal substrates with different mean terrace widths.

\begin{figure}[ht]
  \includegraphics{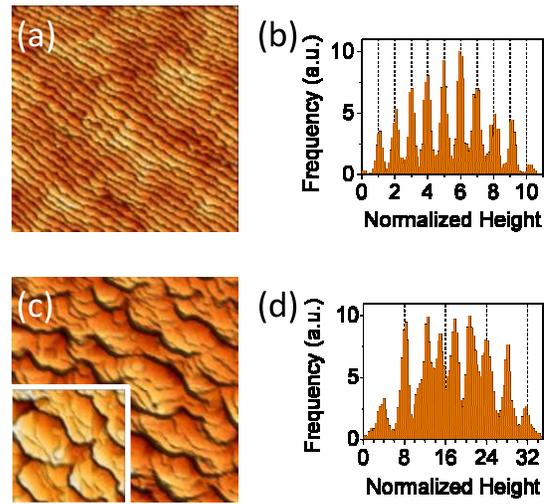}
  \caption{$4 \times 4$ $\mu m^{2}$ AFM topographs of SrRuO$_{3}$ films grown at 610\textdegree C
    in the step flow mode with mean step velocities (a) V = 10.2 nm/s and (c) V = 8.3 nm/s.
    Mean terrace width of SrTiO$_{3}$ substrate is 120 nm (a,b), 150 nm (c,d) and 250 nm
    for inset of (c). Height histograms taken on a $1 \times 1$ $\mu m^{2}$ area of image (a)
    and on the whole image in (c) are displayed in (b) and (d), respectively.
    The height is normalized to the one of a strained SrRuO$_{3}$ unit cell (see text).}
 \label{sro-afm}
 \end{figure}

Substrates of SrTiO$_{3}$ were prepared according to an established procedure \cite{kaw94,bla98}
to ensure their chemical termination to be TiO$_{2}$. The mean terrace width varies from 65 nm
to 370 nm, corresponding to miscut angles from 0.06\textdegree \ to 0.3\textdegree. The substrates
were systematically characterized with atomic force microscopy (AFM) in contact mode prior to their
introduction into the growth chamber, in particular to check the single atomic
termination through the friction contrast. From X-Ray diffraction (XRD) and AFM
mean terrace width values are obtained with a relative accuracy better than 10\%. Thin films of
SrRuO$_{3}$ were grown on these substrates by PLD using a KrF excimer Laser ($\lambda = 248$ nm)
at a growth temperature of 610\textdegree C under 120 mTorr of pure oxygen atmosphere.
Laser energy was set to 200 mJ and the fluence was further refined with an attenuator
to get 3 J/cm$^{2}$ on the target. A homogenizer is used to give the Laser beam a flat-top
profile to avoid droplet formation on the sample surface. To ensure a good control of both
film thickness and mean deposition flux, an \textit{in situ} optical reflectometry technique
is used \cite{est10}. The thickness could thus be followed in real time taking advantage of the
interferences between the light reflected from the film surface and from the film-substrate interface.
XRD measurements were performed to double-check film thickness and to verify that all
SrRuO$_{3}$ films were fully strained on SrTiO$_{3}$. Their unit cell parameter
perpendicular to the surface has been found equal to 0.395 nm, larger than SrRuO$_{3}$ bulk value
of 0.393 nm thus confirming an in-plane compressive strain for the films. Except if indicated otherwise all
films are 8.7 nm thick, corresponding to 22 ML of strained SrRuO$_{3}$. The deposition flux $F$ is
taken as $F=pN_{p}$ with $p$ the Laser repetition rate and $N_{p}$ the (fixed) amount of SrRuO$_{3}$
deposited per pulse, set to 0.055 ML for all depositions. Typical repetition rates were between 0.6 Hz
and 2 Hz, for a flux ranging from $3.3 \times 10^{-2}$ ML/s to $1.1 \times 10^{-1}$ ML/s.

Figure \ref{sro-afm} shows AFM topographs of typical morphologies observed after SrRuO$_{3}$
growth in the stable step flow regime (a,b) and in the step-bunching instability region (c,d).
Height histograms are given in (b,d), with the height normalized to the single-step height of
0.395 nm found for our strained SrRuO$_{3}$ films. While after stable step flow all terraces
are separated by single steps, after 8.7 nm-step flow in the unstable regime step bunches
comprised of 3 to 4 steps are clearly seen in the histogram. Single steps are still
found on the terraces in between the step bunches, which are not straight but curved as seen
on the AFM image. Inset of Fig. \ref{sro-afm}(c) is a $4 \times 4$ $\mu m^{2}$ topograph of a
film grown with the same parameters as for the one of Fig. \ref{sro-afm}(c) but on a
SrTiO$_{3}$ substrate with 250 nm-wide terraces instead of 150 nm. Flux has been adjusted to get
the same step velocity of 8.3 nm/s. A similar bunched morphology is found for both films, with
roughly the same number of steps in the bunches and thus a bunch separation scaling with initial
terrace width.

\begin{figure}[h]
  \includegraphics{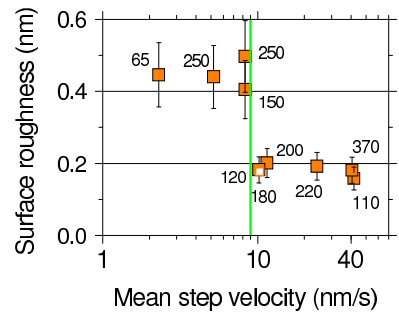}
  \caption{Transition from stable step flow to step bunching as seen through surface roughness. A critical
  step velocity materialized by the vertical line is found at V$^{*}$ = 9 nm/s. Open square data point
  is for a film grown successively at V = 7.7 nm/s and V = 10.9 nm/s, the latter velocity value being the
  one reported in the graph. Each data point is labeled with the mean terrace width (nm)
  of the corresponding SrTiO$_{3}$ substrate.}
 \label{sro-rms}
 \end{figure}

The drastic change between stable step train (Fig. \ref{sro-afm}(a))and unstable step
bunches (Fig. \ref{sro-afm}(c)) is obtained for a rather small increase of deposition flux and
thus of mean step velocity. Moreover, all bunched morphologies are already well
developed after deposition of 8.7 nm of strained SrRuO$_{3}$. These observations point to a sharp
transition from stable to unstable step flow, which was further investigated through varying both
deposition flux and substrate mean terrace width. Noting that a bunched morphology gives an
increased surface roughness, the latter parameter has been chosen to trace the transition. Results
are plotted as a function of mean step velocity in Fig. \ref{sro-rms}, with the root mean square
(rms) roughness measured from $4 \times 4$ $\mu m^{2}$ AFM images. A clear transition from stable
step flow and roughness around 0.2 nm to unstable one and roughness close to 0.4 nm is found
with a critical step velocity V$^{*}$ = 9 nm/s. We checked explicitly that upon increasing the
deposition flux further on substrates with the largest terrace widths, typically above 200 nm,
monolayer-high islands are observed on the terraces with a growth dominated by island nucleation
and coalescence \cite{sflow}.

The origin of the experimentally observed sharp transition can be captured in the
step velocity model accounting for strain effects introduced in Ref. \onlinecite{zha05}. In
the absence of step interactions, step flow can be stabilized against step bunching due to a preferred
incorporation of diffusing adatoms to the ascending step edges. A characteristic length $\ell_{s}$ can
be defined for adatom incorporation to a step from the upper terrace, often traced back to an excess
energy barrier for diffusion over the step edge. The ratio $f_{s} \equiv \ell_{s}/(L+\ell_{s})$
then measures the strength of the incorporation asymmetry,
assuming instantaneous attachment from the lower side of a step edge. In a step train picture, a fraction
$(1+f_{s})/2$ of all the adatoms diffusing on a given terrace will eventually attach to the ascending
step edge, while the remaining $(1-f_{s})/2$ will be incorporated at the descending step, for an effective
adatom current in the step-up direction. Now considering step interactions, it has been shown that for a
strained film there is an attractive interaction between steps in addition to the classical elastic
repulsion, and all strained step trains could thus be thermodynamically unstable against step bunching
\cite{ter95}. Introducing these interactions in a step velocity model, the competition between stabilizing
step-up adatom current and attractive interaction between steps results in a critical step velocity
V$^{*}$ for step flow, below which step bunching occurs \cite{zha05,zha07}:
 
\begin{equation}
	V^{*} = \Omega^{2}\pi^{2}\frac{\alpha_{1}c_{0}D}{k_{B}Tf_{s}L\ell_{s}}
\end{equation}
	
Here $\Omega$ is the area of the surface unit cell, $\alpha_{1}$ is the elastic constant measuring the strength
of the attractive interaction between steps \cite{ter95}, $c_{0}$ the equilibrium concentration of adatoms in
the vicinity of a straigth step, $D$ adatom diffusion coefficient,  and $k_{B}$ the Boltzmann constant.
As factors promoting an increase of V$^{*}$ and thus step bunching, the mass transfer coefficient expressed
through $c_{0}D/k_{B}T$ is found alongside $\alpha_{1}$ given that the driving force for step bunching does
not require actual growth and is present even for vanishing deposition flux. This is at variance with the
term $f_{s}L\ell_{s}$ stemming from the kinetics of the flowing step train with an effective step-up adatom
current. For a weak attachment barrier $f_{s}L\sim\ell_{s}$ and the critical step velocity does not depend on
mean terrace width L, while for moderate or strong barrier V$^{*}$ increases with decreasing L.

Looking at our data, the fact that a transition from unstable to stable step flow can be detected
at V$^{*}$ = 9 nm/s for a wide range of mean terrace widths especially around the transition points
at a small barrier for incorporation at a descending step, i.e. the $\ell_{s}<<L$ case. For example,
the growth of SrRuO$_{3}$ at V = 8.3 nm/s is found unstable for L = 150 nm and L = 250 nm (Fig. \ref{sro-afm}(c)),
while going to the smaller terrace width L = 120 nm the step flow is already stabilized at V = 10.2 nm/s
(Fig. \ref{sro-afm}(a)). Still an asymetry of incorporation between both sides of a step has to
be invoked otherwise the step train would be always unstable with respect to step bunching \cite{ter95}.

\begin{figure}[h]
  \includegraphics{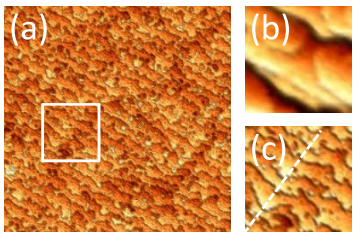}
  \caption{Transition \textit{during growth} from step bunching at V = 7.7 nm/s (b) to stable step flow
  at V = 10.9 nm/s (a,c) through the tuning of the mean deposition flux by adjusting the Laser
  repetition rate. AFM topographs are $4 \times 4$ $\mu m^{2}$ in (a) with close-ups of
  $1 \times 1$ $\mu m^{2}$ in (b,c). Image shown in (b) is for a film for which growth as been stopped
  in the step bunching regime, to be compared with (c). The dotted line in (c) is a guide
  to the eye. Mean terrace width of substrate is 180 nm.}
 \label{sro-debunch}
 \end{figure}

In order to demonstrate the ability to pass from unstable to stable step flow through the tuning
of step velocity, first a SrRuO$_{3}$ layer was grown in the step bunching regime up to 17 nm thickness
with V = 7.7 nm/s, followed by 48 nm of stable step flow at V = 10.9 nm/s. The mean terrace width
of the SrTiO$_{3}$ substrate was 180 nm, and the deposition flux $F$ was changed by tuning the
Laser repetition rate. As seen in Figure \ref{sro-debunch}, final morphology is a smooth step train
with only single steps. In several places an in-phase meandering of steps can be observed
which is a signature of a flowing step train with effective step-up adatom current. For example the close-up
in Fig. \ref{sro-debunch}(c) shows step edge profiles of neighboring steps aligned along the
dotted line. According to Ref. \onlinecite{zha07}, step bunching only occurs above a critical
thickness which increases with terrace width, so that a transient stability region is found for step flow
up to this thickness. As all the bunched morphologies are here well developped already for 8.7 nm deposited
SrRuO$_{3}$, the critical thickness should be lower than this value. We thus believe the step flow growth
observed for $V > V^{*}$ to be outside the transient region. Added support for a persistent step flow is given
by the fact that a bunched morphology is fully turned into a step train with equidistant steps upon
growth at $V > V^{*}$ (Fig. \ref{sro-debunch}), which would not be expected if step flow was ultimately
unstable in this step velocity region. It should be noted that the instability does not initially proceed
through the bunching of straigth steps as step meandering is seen on all our bunched films
(see Fig. \ref{sro-afm}(c)). While the interplay between step bunching and step meandering is non
trivial to address theoretically in particular as for the attractive interaction between steps
\cite{ter03,tho07}, such morphologies have already been reported for the step flow growth of metals
on vicinal surfaces \cite{cu03}.

In conclusion, we demonstrated the ability to tune the growth mode of SrRuO$_{3}$ on (001) SrTiO$_{3}$
from stable to unstable step flow with formation of step bunches through the control of the mean
step velocity. This parameter sums up the effects of deposition flux (Laser repetition rate) and
substrate mean terrace width, stressing the importance of the latter and thus of substrate preparation.
A critical step velocity V$^{*}$ = 9 nm/s below which step bunching occurs was found at 610\textdegree C and
is independent of mean terrace width. V$^{*}$ is expected to increase with increasing temperature as
the adatom incorporation asymmetry which stabilizes the step train against step bunching is a decreasing
effect with temperature. As the repetition rate and thus the mean step velocity can be adjusted during
growth it opens the possibility for \textit{in situ} growth mode manipulation in order to control the film step
structure, the latter being monitored for example with grazing incidence electron diffraction or optical
reflectometry techniques. This is of particular interest for the elaboration of oxide superlattices where
maintaining a stable step flow going from one layer to the next through a proper tuning of
deposition flux is one possible way to unsure smooth interfaces between layers.

This work has been supported by the Agence Nationale pour la Recherche (MINOS project 2008-2011).


\begin{thebibliography}{0}
\bibitem{sflow} In the case of PLD growth, a transient supersaturation of hyperthermal species coming
                from the target is created on the surface during each Laser pulse. Step flow is achieved
                when all deposited atoms are eventually incorporated at pre-existing steps. When stable
                immobile islands are formed on the terraces, growth proceeds through island nucleation
                and coalescence possibly mixed with step flow depending on terrace width and deposition
                parameters. Such islands can be detected with AFM after deposition, provided
                the deposition is not stopped exactly at layer completion in the case of perfect layer-by-layer
                growth.
\end{thebibliography}
\end{document}